\documentclass[twocolumn,showpacs,preprintnumbers,amsmath,amssymb,floatfix,prl]{revtex4}
\usepackage{epsfig}
\begin{document}
\title{Evidence for polarization of gluons in the proton}
\author{Daniel de Florian}  
\email{deflo@df.uba.ar}
\author{Rodolfo Sassot} 
\email{sassot@df.uba.ar}
\affiliation{Departamento de F\'{\i}sica and IFIBA, Facultad de Ciencias Exactas y Naturales, 
Universidad de Buenos Aires, Ciudad Universitaria, Pabell\'on\ 1 (1428) Buenos Aires, 
Argentina}
\author{Marco Stratmann}  
\email{marco.stratmann@uni-tuebingen.de}
\affiliation{Institute for Theoretical Physics, T\"ubingen University, Auf der Morgenstelle 
14, 72076 T\"ubingen, Germany}
\affiliation{Physics Department, Brookhaven National Laboratory, Upton, NY~11973, USA}
\author{Werner Vogelsang} 
\email{werner.vogelsang@uni-tuebingen.de}
\affiliation{Institute for Theoretical Physics, T\"ubingen University, Auf der Morgenstelle 
14, 72076 T\"ubingen, Germany}

\begin{abstract}
We discuss the impact of recent high-statistics RHIC data on the determination of the gluon polarization 
in the proton in the context of a global QCD analysis of polarized parton distributions. We find clear evidence
for a non-vanishing polarization of gluons in the region of momentum fraction and at the scales 
mostly probed by the data. Although information from low momentum fractions is presently lacking, 
this finding is suggestive of a significant contribution of gluon spin to the proton spin, thereby limiting 
the amount of orbital angular momentum required to balance the proton spin budget. 
\end{abstract}


\maketitle


{\it Introduction.---} The gluon helicity distribution function $\Delta g(x)$ of the proton has long
been recognized as a fundamental quantity characterizing the inner structure of the nucleon. 
In particular, its integral $\Delta G\equiv\int_0^1dx \Delta g(x)$ over all gluon momentum fractions $x$ 
may in $A^+=0$ light-cone gauge be interpreted as the gluon spin contribution to the proton 
spin~\cite{Leader:2013jra}. As such, $\Delta G$ is a key ingredient to the proton 
helicity sum rule,
\begin{equation}
\frac{1}{2}=\frac{1}{2}\Delta \Sigma + \Delta G+L_q+L_g \;, \label{ssr}
\end{equation}
where $\Delta \Sigma$ denotes the combined quark and antiquark spin contribution
and $L_{q,g}$ are the quark and gluon orbital angular momentum contributions. For 
simplicity, we have omitted the renormalization scale $Q$ and scheme dependence of all quantities.

It is well known that the quark and gluon helicity distributions can be probed
in high-energy scattering processes with polarized nucleons, allowing access
to $\Delta \Sigma$ and $\Delta G$. Experiments on polarized 
deep inelastic lepton-nucleon scattering (DIS) performed since the late 
eighties~\cite{Aidala:2012mv} have shown that relatively little of the proton spin 
is carried by the quark and antiquark spins, with a typical value $\Delta \Sigma\sim 
0.25$~\cite{Aidala:2012mv,ref:dssv,ref:others}. 
The inclusive DIS measurements have, however, very little
sensitivity to gluons. Instead, the best probes of $\Delta g$ are offered by 
polarized proton-proton collisions available at the BNL Relativistic Heavy Ion Collider 
(RHIC)~\cite{Aschenauer:2013woa}. Several processes in $pp$ collisions, 
in particular jet or hadron production at high transverse momentum $p_T$, receive 
substantial contributions from gluon-induced hard scattering, hence opening a window on
$\Delta g$ when polarized proton beams are used.

The first round of results produced by RHIC until 2008~\cite{Aschenauer:2013woa}
were combined with data from inclusive and semi-inclusive DIS in a
next-to-leading order (NLO) global QCD analysis~\cite{ref:dssv}, hereafter
referred to as ``DSSV analysis''. 
One of the main results of that analysis was that the RHIC data -- within their 
uncertainties at the time -- did not show any evidence of a polarization of gluons
inside the proton. In fact, the integral of $\Delta g$ over the region $0.05\leq x\leq
0.2$ of momentum fraction primarily accessed by the RHIC experiments was found to be very 
close to zero. Other recent analyses of nucleon spin structure~\cite{ref:others} did not fully include RHIC data; as a result $\Delta g$
was left largely unconstrained.

Since the analysis~\cite{ref:dssv}, the data from RHIC have vastly improved.
New results from the 2009 run~\cite{pibero,Adare:2014hsq} at center-of-mass 
energy $\sqrt{s}=200$~GeV have significantly 
smaller errors across the range of measured $p_T$. This will 
naturally put tighter constraints on $\Delta g(x)$ and may extend the range of $x$ 
over which meaningful constraints can be obtained. A striking
feature is that the {\sc{Star}} jet data~\cite{pibero} now exhibit a
double-spin asymmetry $A_{LL}$ that is clearly non-vanishing over the whole range
$5 \lesssim p_T \lesssim 30~\mathrm{GeV}$, in contrast to the previous results. Keeping
in mind that in this regime jets are primarily produced by gluon-gluon and 
quark-gluon scattering, this immediately suggests that gluons inside the proton 
might be polarized. At the same time, new {\sc{Phenix}} data for $\pi^0$ production~\cite{Adare:2014hsq}
still do not show any significant asymmetry, and it is of course important to 
reveal whether the two data sets provide compatible information.  
In this letter, we assess the impact of the 2009 RHIC data sets on $\Delta g$
in the context of a new NLO global analysis of helicity parton densities.

{\it Global analysis and new and updated data sets.---} 
As just described, the key ingredients to our new QCD analysis are the 2009 
{\sc{Star}}~\cite{pibero} and {\sc{Phenix}}~\cite{Adare:2014hsq} data on the 
double-spin asymmetries for inclusive jet and $\pi^0$ production. At the same
time, we also update some of the earlier RHIC results used in \cite{ref:dssv} and
add some new DIS data sets by the {\sc{Compass}} experiment. More specifically, we now
utilize the final {\sc{Phenix}} $\pi^0$ data from run-6 at $\sqrt{s}=200$~GeV~\cite{Adare:2008aa}
and 62.4~GeV~\cite{Adare:2008qb}, the final {\sc{Star}} jet results
from run-5 and run-6~\cite{Adamczyk:2012qj}, and
the recent inclusive~\cite{Alekseev:2010hc} and 
semi-inclusive~\cite{Alekseev:2010ub} DIS data sets from {\sc{Compass}}. 
As far as the impact on 
$\Delta g$ is concerned, the data sets~\cite{pibero,Adare:2014hsq} clearly dominate. The 
{\sc{Compass}} data sets will primarily affect the quark and antiquark
helicity distributions as reported in~\cite{deFlorian:2011cr}.

The method for our global analysis has been described in detail in~\cite{ref:dssv} 
and will not be presented here again. It is based on an efficient Mellin-moment technique 
that allows one to tabulate and store the computationally most demanding parts of a 
NLO calculation prior to the actual analysis. In this way, the 
evaluation of the relevant spin-dependent $pp$ cross sections \cite{ref:ppxsec}
becomes so fast that it can be easily performed
inside a standard $\chi^2$ minimization analysis. 
As a small technical point,
we note that {\sc{Star}} has moved to the ``anti-$k_t$'' jet algorithm~\cite{Cacciari:2008gp} 
for their analysis of the data from the 2009 run. In order to match this feature,
we use the NLO expressions derived in~\cite{Mukherjee:2012uz} for the
polarized case. 
As in our previous DSSV analysis~\cite{ref:dssv}, standard Lagrange multiplier (L.M.)
and Hessian techniques are employed in order to assess the uncertainties
of the polarized parton distributions determined in the fit.

We adopt the same flexible functional form as in~\cite{ref:dssv}
to parametrize the NLO helicity parton densities
at the initial scale $Q_0=1$~GeV, for instance, 
\begin{equation}
x\Delta g(x,Q_0^2)=N_g x^{\alpha_g}(1-x)^{\beta_g}\left(1+\eta_g x^{\kappa_g}\right)\, ,
\label{dginp}
\end{equation}
with free parameters $N_g,\alpha_g,\beta_g,\eta_g$, and $\kappa_g$. Note that this parameterization 
allows for a node in the distribution, as realized by the central gluon density of the DSSV
analysis~\cite{ref:dssv}. We enforce positivity 
$|\Delta f|/f\leq 1$ of the parton densities, using the unpolarized distributions $f(x,Q^2)$
of~\cite{Martin:2009iq}, from where we also adopt the running of the strong coupling. 
We use the same set for computing the spin-averaged cross sections in the denominators
of the spin asymmetries. 

\begin{figure}[t]
\vspace*{-0.7cm}
\epsfig{figure=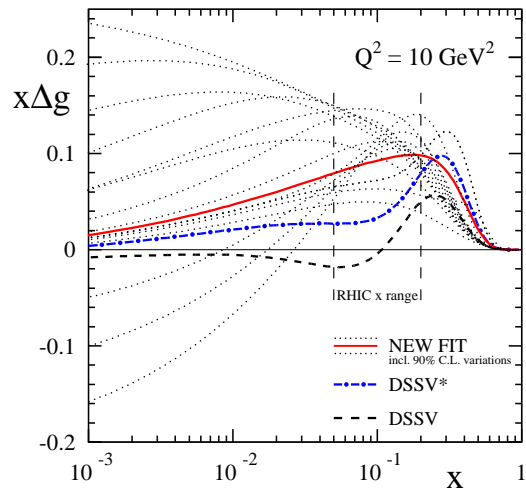,width=0.45\textwidth}
\vspace*{-0.7cm}
\caption{Gluon helicity distribution at $Q^2=10\,\mathrm{GeV}^2$
for the new fit, the original DSSV 
analysis of~\cite{ref:dssv}, and for an updated analysis without using the
new 2009 RHIC data sets (DSSV*, see text). The dotted lines present 
the gluon densities for alternative fits that are within the $90\%$ C.L.\ limit. 
The $x$-range primarily probed by the RHIC data is indicated by the two
vertical dashed lines.
\label{fig:Deltag}}
\vspace*{-3mm}
\end{figure}

{\it Results of global analysis.---} Figure~\ref{fig:Deltag} shows our new result for
the gluon helicity distribution $\Delta g(x,Q^2)$ at $Q^2=10$~GeV$^2$. The solid
line presents the updated central fit result, with the dotted lines corresponding to additional fits 
that are within the $90\%$ confidence level (C.L.) interval. In defining 
this interval, we follow the strategy adopted in Ref.~\cite{Martin:2009iq}.
These alternative fits may be thought of as spanning
an uncertainty band around $\Delta g$ within this tolerance and for the adopted functional form (\ref{dginp}).
The dot-dashed curve represents
the result of a fit -- henceforth labelled as ``DSSV*'' -- for which we only include the updates to the various RHIC
data sets already used for the original DSSV analysis \cite{ref:dssv} (dashed line), i.e., we exclude all
the new 2009 data \cite{pibero,Adare:2014hsq}. The new {\sc{Compass}} inclusive~\cite{Alekseev:2010hc} and 
semi-inclusive~\cite{Alekseev:2010ub} DIS data sets have little impact on $\Delta g$ and are
included in the DSSV* fit.
%

The striking feature of our new polarized gluon distribution is its much larger size as compared 
to the one of the DSSV analysis~\cite{ref:dssv}. For $Q^2=10$~GeV$^2$, it is positive throughout 
and clearly away from zero in the regime $0.05\leq x\leq 0.2$ predominantly probed by the RHIC
data, as is demonstrated by the alternative fits spanning the 90\% C.L.\ interval. 
In contrast to the original DSSV gluon distribution, the new $\Delta g$ does not 
show any indication of a node in the RHIC $x$-range~\footnote{Our new $\Delta g$ in Eq.~(\ref{dginp}) 
is given by $N_g=1774$, $\alpha_g=5.6$, $\beta_g=9.0$, $\eta_g=0.0023$, 
and $\kappa_g=-3.0$.
A {\sc Fortran} code of our new DSSV set is available upon request.
}.
It is interesting to notice
that the DSSV* fit, without the new 2009 but with updated earlier RHIC data sets, 
already tends to have a positive $\Delta g$. This trend is then very much strengthened, 
in particular, by the 2009 {\sc{Star}} data \cite{pibero}.

\begin{figure}[t]
\vspace*{-0.6cm}
\epsfig{figure=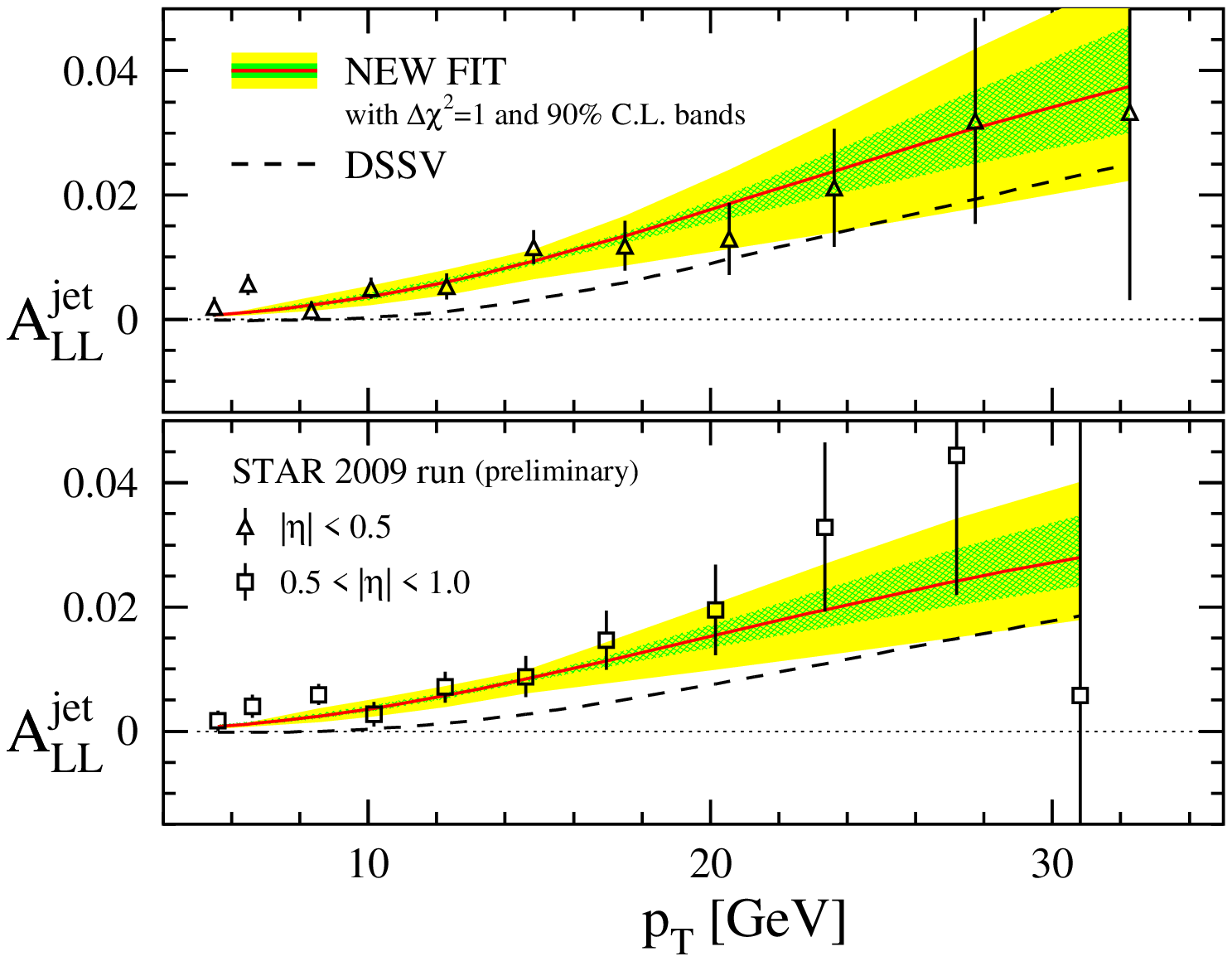,width=0.45\textwidth}
\vspace*{-0.3cm}
\caption{Latest preliminary {\sc{Star}} data~\cite{pibero} for the double-spin asymmetry in jet
production for two rapidity ranges
compared to the results of our new and original \cite{ref:dssv} analyses. 
The inner and outer bands correspond to $\Delta\chi^2 = 1$ 
and 90 \% C.L., respectively.
\label{fig:rhic_jets}}
%
\vspace*{-3mm}
\epsfig{figure=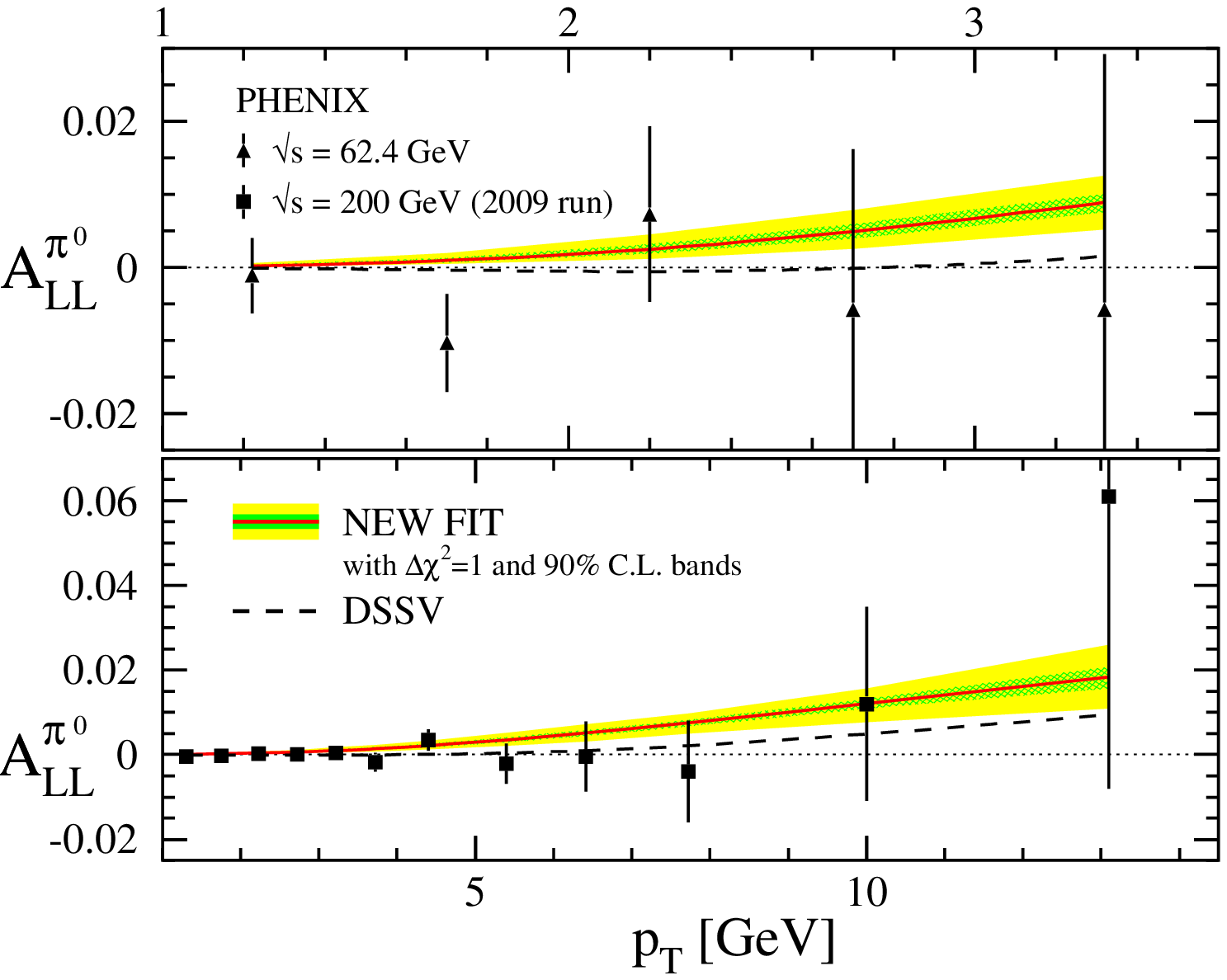,width=0.45\textwidth}
\vspace*{-0.3cm}
\caption{As Fig.~\ref{fig:rhic_jets}, but comparing to the {\sc{Phenix}} data~\cite{Adare:2014hsq,Adare:2008qb} 
for the double-spin asymmetry in $\pi^0$ production at $\sqrt{s}=62.4$~GeV (upper panel)
and $\sqrt{s}=200$~GeV (lower panel).
\vspace*{-3mm}
\label{fig:rhic_pi0}}
\end{figure}

Figure~\ref{fig:rhic_jets} shows the comparison to 
the new {\sc{Star}} jet data~\cite{pibero} obtained with our new set of 
spin-dependent distributions. As in the analysis itself, we have chosen both the 
factorization and renormalization scales as $p_T$. 
{\sc{Star}} presents results for two rapidity ranges,
$|\eta|<0.5$ and $0.5<|\eta|<1$.
It is evident that the new fit describes the data very well in both ranges.
We also illustrate the uncertainties corresponding to our analysis, using
the L.M.\ method with a tolerance $\Delta\chi^2=1$ (inner bands)
and 90 \% C.L. (outer bands). Also shown is the result
for our previous DSSV analysis~\cite{ref:dssv}. As one can see, it falls considerably
short of the data in the region $10 \lesssim p_T \lesssim 20\,\mathrm{GeV}$, 
where it barely touches the new uncertainty band. This precisely demonstrates the fact mentioned
earlier that the 2009 jet data \cite{pibero} tend to exhibit a somewhat higher asymmetry than 
previously, resulting in larger gluon polarization in the new fit. 
Comparing to the results of~\cite{ref:dssv} one finds that the uncertainty 
bands for our new fit have become significantly narrower than for the DSSV one. 
The new analysis, including updates and new data, is within the uncertainty estimate for the old 
DSSV fit~\cite{ref:dssv}. 
%

Figure~\ref{fig:rhic_pi0} shows corresponding comparisons to the {\sc{Phenix}}
data for $A_{LL}$ in $\pi^0$ production at $\sqrt{s}=62.4\,\mathrm{GeV}$ and 
$200\,\mathrm{GeV}$ \cite{Adare:2014hsq,Adare:2008qb}. In contrast to jet production, 
the asymmetries are consistent with zero within uncertainties. As a result, they are still perfectly 
described by a calculation based on the original DSSV analysis with its small gluon polarization
exhibiting a node. 
Within our new analysis, we obtain also here a larger spin asymmetry that still
describes the data very well. In this sense, the new {\sc{Star}} and {\sc{Phenix}} data
sets are mutually consistent. 

\begin{figure}[t]
\vspace*{-0.7cm}
\epsfig{figure=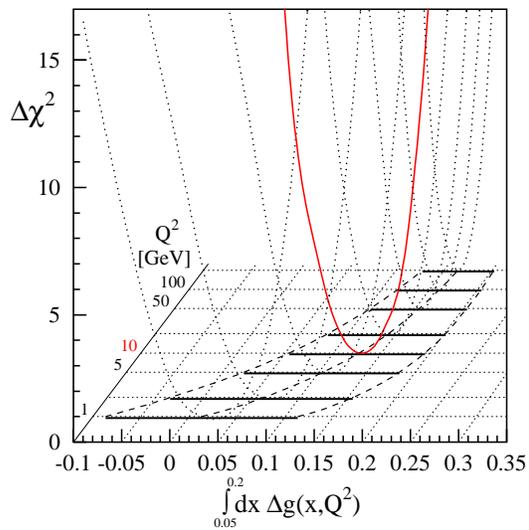,width=0.45\textwidth}
\vspace*{-0.1cm}
\caption{Change of the $\Delta \chi^2$ profile of the truncated first 
moment of $\Delta g$ in the RHIC $x$-range with $Q^2$. The solid lines at the
base of the plot indicate the $90\%$ C.L. interval.
\label{fig:Q2}}
\vspace*{-3mm}
\end{figure}

It is worth pointing out in this context that the RHIC jet and pion data sets probe $\Delta g(x)$ at 
different scales $Q$, owing to the different ranges in transverse momentum accessed. As a result, 
the scale evolution of $\Delta g(x)$ plays a role here, a point that we will elaborate on now.
Figure~\ref{fig:Q2} shows the variation of the total $\chi^2$ of the fit
as a function of the truncated first moment in the RHIC $x$-range, 
$\int_{0.05}^{0.2}dx \Delta g(x,Q^2)$, for various values of $Q^2$. 
The solid curve corresponds to $Q^2=10$~GeV$^2$,
which once more demonstrates that the truncated moment is clearly positive at this scale
within our estimated $90\%$ C.L.\ variations indicated at the base of the plot.
As one can see, towards lower scales the central value of the moment
decreases and its uncertainty increases. 
The observed relatively strong scale dependence
is reminiscent of that known for the full first moment~\cite{Altarelli:1988nr}.
It is a significant factor for the consistency between the {\sc{Phenix}} $\pi^0$ data taken at
lower scales and the {\sc{Star}} jet data at higher scales. In the original DSSV analysis \cite{ref:dssv}
the $\chi^2$ was dominated by data at lower scales, hence resulting in the nearly
vanishing $\Delta g$. A feature related to these observations is that
our new gluon distribution is peaked at relatively high $x$ at
the input scale, in fact just above the RHIC region. Evolution 
then pushes the distribution toward lower $x$, making it compatible
with the {\sc{Star}} data which probe it at much higher scales.
%

Ultimately, one is interested, of course, in a reliable determination of the full integral $\Delta G$
entering in (\ref{ssr}).
RHIC data mainly probe the region $0.05\leq x\leq 0.2$, but the more precise 2009 results help to constrain
$\Delta g(x)$ better down to somewhat lower values $x\simeq 0.02$.
Here, some very limited information on $\Delta g$ is also available from scaling violations of the DIS structure
function $g_1$ which is, of course, fully included in our global QCD analysis.
Overall, the constraints on $\Delta g(x)$ in, say, the regime $0.001\leq x\leq 0.05$ are much
weaker than those in the RHIC region, as can be inferred from Fig.~\ref{fig:Deltag}.
Very little contribution to $\Delta G$ is expected to come from $x>0.2$. 

\begin{figure}[t]
\vspace*{-0.7cm}
\epsfig{figure=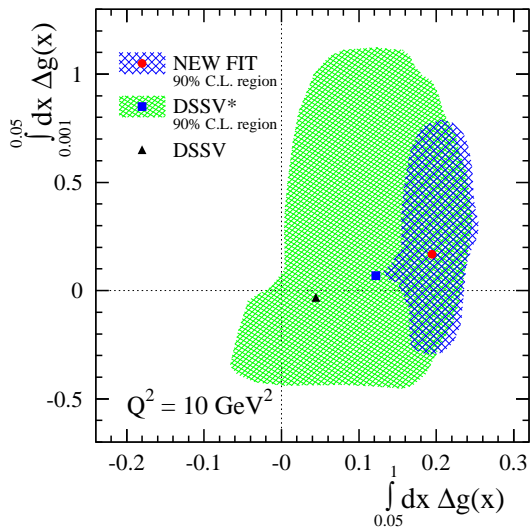,width=0.45\textwidth}
\vspace*{-0.3cm}
\caption{$90\%$ C.L.\ areas in the plane spanned by the truncated moments of 
$\Delta g$ computed for $0.05\leq x\leq 1$ and $0.001\leq x\leq 0.05$ at $Q^2=10\,\mathrm{GeV}^2$.
Results for DSSV, DSSV*, and our new analysis, with the symbols 
corresponding the respective values of each central fit, are shown.
\label{fig:truncg}}
\vspace*{-4mm}
\end{figure}

Figure~\ref{fig:truncg} shows our estimates for the $90\%$ C.L.\ area 
in the plane spanned by the truncated moments of $\Delta g$ calculated in 
$0.05\leq x\leq 1$ and $0.001\leq x\leq 0.05$ for $Q^2=10\,\mathrm{GeV}^2$.
Results are presented both for the DSSV* and our new fit.
The symbols in Fig.~\ref{fig:truncg} denote the
actual values for the best fits in the DSSV, DSSV*, and the present analyses. 
We note that for our new central fit the combined integral $\int_{0.001}^1 dx \Delta g(x,Q^2)$ accounts 
for over $90\%$ of the full $\Delta G$ at $Q^2=10\,\mathrm{GeV}^2$.
Not surprisingly, the main improvement in our new analysis is to shrink the allowed
area in the horizontal direction, corresponding to the much better determination of $\Delta g(x)$
in range $0.05\leq x\leq 0.2$ by the 2009 RHIC data.
Evidently, the uncertainty
in the smaller-$x$ range is still very significant, and better small-$x$ probes
are badly needed. Data from the 2013 RHIC run at $\sqrt{s}=510\,\mathrm{GeV}$ 
may help here a bit. In the future, an Electron Ion Collider would provide
the missing information, thanks to its large kinematic reach in $x$ and $Q^2$~\cite{ref:eic}.
%

{\it Conclusions and outlook.---} We have presented a new global analysis of helicity
parton distributions, taking into account new and updated experimental results.
In particular, we have investigated the impact of the new data on 
$A_{LL}$ in jet and $\pi^0$ production from RHIC's 2009 run. 
For the first time, we find that the jet data clearly imply a polarization
of gluons in the proton at intermediate momentum scales, in the 
region of momentum fractions accessible at RHIC. This constitutes
a new ingredient to our picture of the nucleon. While it is too early to 
draw any reliable conclusions on the full gluon spin contribution to the 
proton spin, our analysis clearly suggests that gluons could contribute
significantly after all. This in turn also sheds a new light on the possible
size of orbital angular momenta of quarks and gluons. We hope that
future experimental studies, as well as lattice-QCD computations that now
appear feasible~\cite{Hatta:2013gta},
will provide further information on $\Delta g(x)$ and eventually 
clarify its role for the proton spin. We plan to present a
full new global analysis with details on all polarized parton
distributions once the 2009 RHIC data have become final and
additional information on the quark and antiquark helicity distributions, in 
particular from final data on $W$ boson production at RHIC, has become available.
Also, on the theoretical side, a new study of pion and kaon fragmentation
functions should precede the next global analysis of 
polarized parton distributions.

\acknowledgments
We thank E.C.\ Aschenauer, K.\ Boyle, P.\ Djawotho, and C.\ Gagliardi 
for useful communications.
M.S.\ was supported in part by U.S. DoE (contract number DE-AC02-98CH10886)
and BNL-LDRD project 12-034. 
This work was supported by CONICET, ANPCyT and UBACyT.        


\end{document}